\documentstyle[psfig]{mn}

\title{The relation between extended radio and line emission for
radio-loud quasars }
\author[X. Cao and D. R. Jiang]
       {Xinwu Cao$^{1,2}$ and D. R. Jiang$^1$ \\
1. Shanghai Astronomical Observatory, Chinese Academy of Sciences, Shanghai,
200030, China \\and National Astronomical Observatories, Chinese Academy
of Sciences, China, cxw@center.shao.ac.cn\\
2. Beijing Astrophysical Center(BAC), Beijing, China}
           
\date{Accepted . Received ; in original form}

\pagerange{\pageref{firstpage}--\pageref{lastpage}}
\pubyear{}

\begin{document}
\maketitle
\label{firstpage}

\begin{abstract}
We explore the relationship between the extended radio and line
emission for a
radio-loud quasar sample including both core-dominated and lobe-dominated
quasars. 
A strong correlation is present between the extended radio and broad-line
emission. The core emission is also correlated with the broad-line 
emission for core-dominated quasars in the sample. The statistic 
behaviour on the core emission of lobe-dominated quasars is rather 
different from that of 
core-dominated quasars. The extended radio luminosity is a good
tracer for jet power, while the core luminosity can only be a jet
power tracer for core-dominated quasars.  

\end{abstract}


\section{Introduction}

The current most favoured models of powering active galactic nuclei (AGNs)
involve gas accretion onto a massive black hole, though the details
are still unclear. Relativistic jets have been observed in many radio-loud
AGNs and are believed to be formed very close to the black holes. In some
theoretical models of the formation of the jet, the power is
generated through accretion and then extracted from the disc/black
hole rotational energy and converted into the kinetic power of the
jet (Blandford \& Znajek; Blandford \& Payne 1982). Recently, numerical
smulations show that the jet can be accelerated from the disk region
very close to the black hole (Koide
et al. 1999). The jet-disc connection has been investigated by many
workers (Rawlings \& Saunders 1991; Falcke \& Biermann 1995; 
Xu \& Livio 1999). An 
effective approach to study the link between these two
phenomena is to explore the relationship between the correponding
emission. A strong correlation is found between the low-frequency radio
and narrow line luminosities of 3C radio sources (Baum \& Heckman 1989;
Rawlings et al. 1989; McCarthy 1993; Tadhunter et al. 1998). The bulk
kinetic power in the jet $Q_{\rm jet}$ can be inferred from its
low-frequency radio luminosity. 
Rawlings \& Saunders (1991) presented
a correlation between $Q_{\rm jet}$ and the narrow line luminosity
$L_{NLR}$. A correlation between the optical and low-frequency
radio luminosities has been confirmed for a sample of steep-spectrum
quasars (Serjeant et al. 1998; Willott et al. 1998). which seems
to argue against the narrow line emission being controlled mainly by
the environment (Dunlop \& Peacock 1993).

The different angles between the radio jet axis and the line of sight
can alter the observational phenomena dramatically. Radio galaxies and
radio-loud quasars are believed to be the same objects viewed at
different angles to the jet axis (Scheuer 1987; Barthel 1989; Antonucci
1993). A dusty torus perpendicular to the jet axis and can 
obscure the central
emission region if the angle between the jet axis and the line of sight is
large enough. The narrow line region in quasars is extended beyond
the putative dusty torus, so that it is believed to be
independent of the jet axis and is therefore isotropic. 

Cao \& Jiang (1999) have performed statistic analyses on the
correlation between the total radio and broad-line emission 
for a sample of radio-loud quasars. 
The broad-line region is ionized by the central source and is 
therefore a good indicator of emission
from the accretion disc. The broad-line data are usually easily available
compared with the narrow line data which is available only for the
sources within a restricted redshift range. 
VLA observations can separate the extended radio
emission from the core emission. In this work, we use Cao \& Jiang's 
sample and collect all available data of extended radio emission 
from the literature to explore the relation between extended radio
emission and broad-line emission. 
 
We describe the adopted
sample in next section. Sections 3 and 4 contain the results and a 
discussion. The cosmological parameters $H_{0}=50$ kms$^{-1}$ Mpc$^{-1}$
and $q_{0}=$0.5 have been adopted in this work.

\section{The  sample}

We start with the sample of Cao \& Jiang (1999), since the broad-line data
of their sample have been well compiled. Their sample is a combination of all
quasars and BL Lac objects with available line data in 1 Jy, S4 and S5
catalogues. The radio flux density limit of S5 catalogue at 5 GHz is 0.25
Jy. There are 198 sources in their sample including 184 quasars and 14 BL Lac
objects. Their sample is drawn from the parent flux-limited sample
selected at 5 GHz. The selection effects may be introduced, i.e., 
those sources without published line flux measurements are 
likely to be biassed towards those with weak lines. 
We believe it would not affect the main results of present 
investigation. We search the literature and collect all available data of VLA
observations on the sources, and find 141 sources with both core and extended
emission data including 128  quasars and  13  BL Lac objects. 
The core and extended flux density is
K-corrected to 5 GHz in the rest frame of the source assuming
$\alpha_{\rm c}=0$ and $\alpha_{\rm e}=-1$ ($f_{\rm core}\propto
\nu^{\alpha_{\rm c}}$, $f_{\rm ext}\propto \nu^{\alpha_{\rm e}}$).
The core and extended
radio flux density at 5 GHz in the rest frame of the sources are listed in
Table 1.  
The ratio $R$ of the core to extended radio luminosity in the rest
frame of the sources is then available
(see column (6) of Table 1). The flux of narrow line [O\,{\sc iii}] 
is also listed in Table 1, since it is the most
frequently observed narrow line for the sources in present sample. 
The total broad-line flux is estimated
by Cao \& Jiang (1999) using the line ratios reported by Francis et al.
(1991).

\begin{figure}
\centerline{\psfig{figure=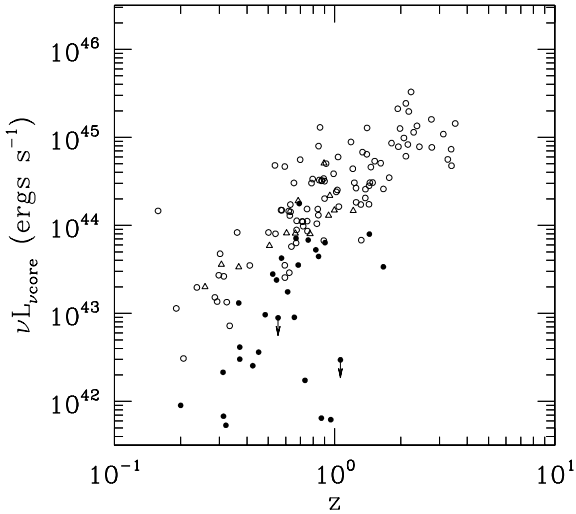,width=8.0cm,height=8.0cm}}
\caption{The radio core luminosity, redshift plane for the sample. The
open circles represent core-dominated quasars, and the full
circles represent  lobe-dominated quasars, while the triangles
represent BL Lac objects.}
\end{figure}

\begin{figure}
\centerline{\psfig{figure=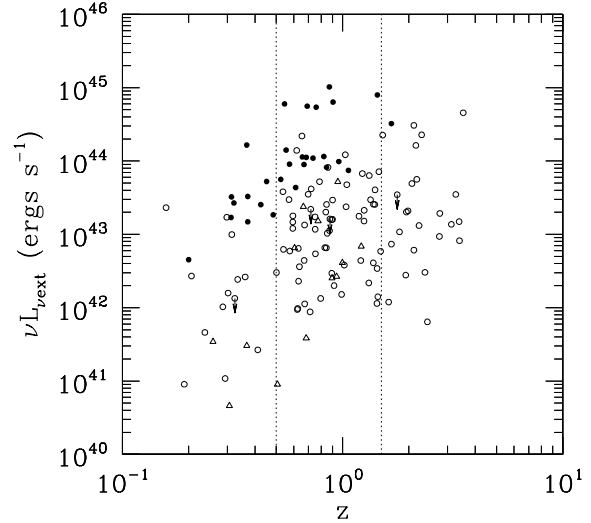,width=8.0cm,height=8.0cm}}
\caption{
Same as Fig. 1, but for the extended radio luminosity. The 
restricted redshift range: $0.5~<z~<1.5$, is indicated by 
the dotted lines. 
}
\end{figure}

\begin{figure}
\centerline{\psfig{figure=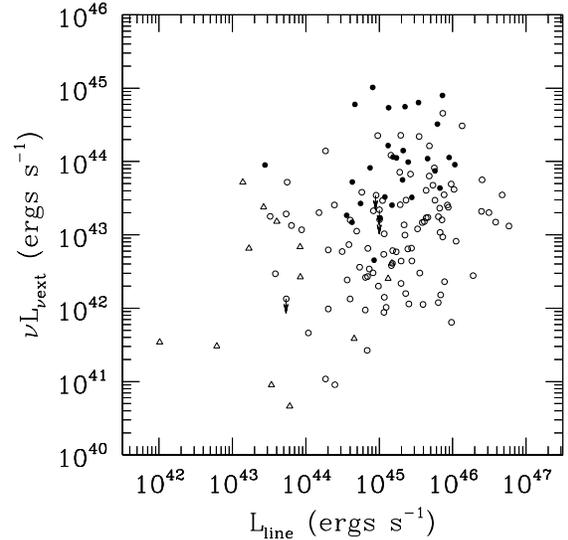,width=8.0cm,height=8.0cm}}
\caption{The extended radio and broad-line luminosity relation
(Symbols as in Fig. 1).
}
\end{figure}

\begin{figure}
\centerline{\psfig{figure=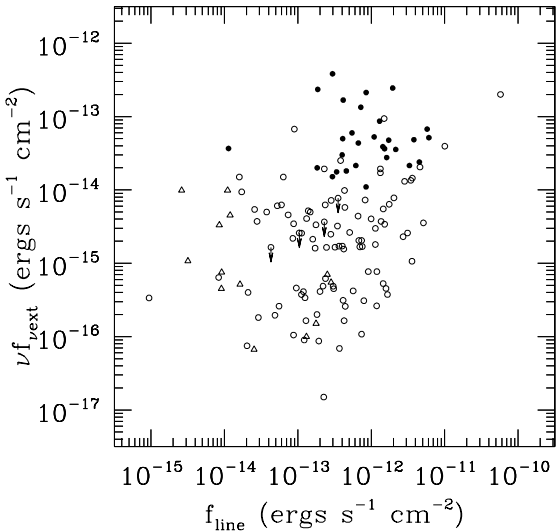,width=8.0cm,height=8.0cm}}
\caption{The extended radio and broad-line flux relation
(Symbols as in Fig. 1).
}
\end{figure}

\begin{figure}
\centerline{\psfig{figure=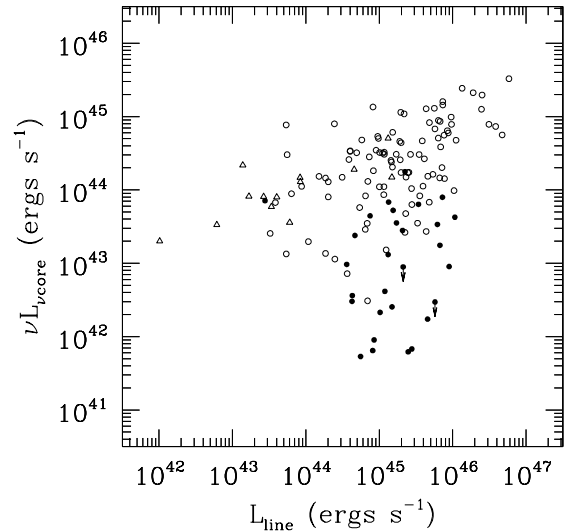,width=8.0cm,height=8.0cm}}
\caption{The radio core and broad-line luminosity relation
(Symbols as in Fig. 1).
}
\end{figure}

\begin{figure}
\centerline{\psfig{figure=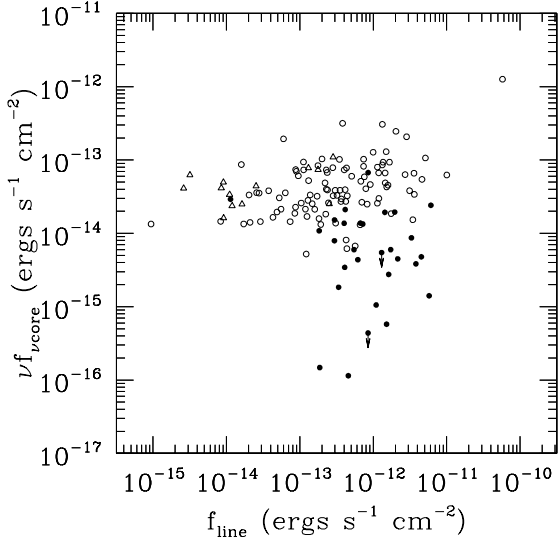,width=8.0cm,height=8.0cm}}
\caption{The radio core and broad-line flux relation
(Symbols as in Fig. 1).
}
\end{figure}

\begin{figure}
\centerline{\psfig{figure=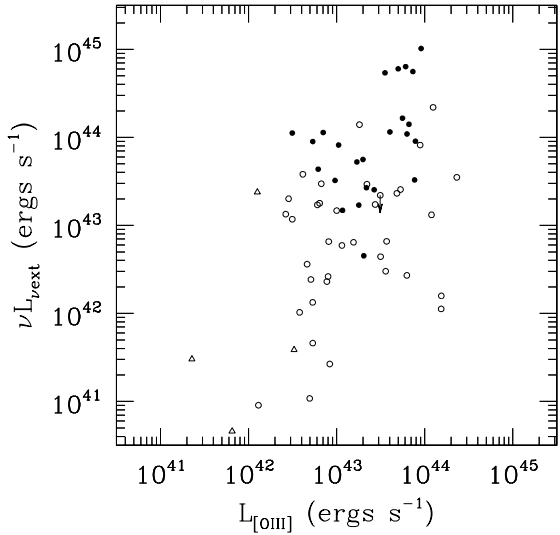,width=8.0cm,height=8.0cm}}
\caption{The extended radio and narrow line [O\,{\sc iii}] luminosity relation
(Symbols as in Fig. 1).
}
\end{figure}

\begin{figure}
\centerline{\psfig{figure=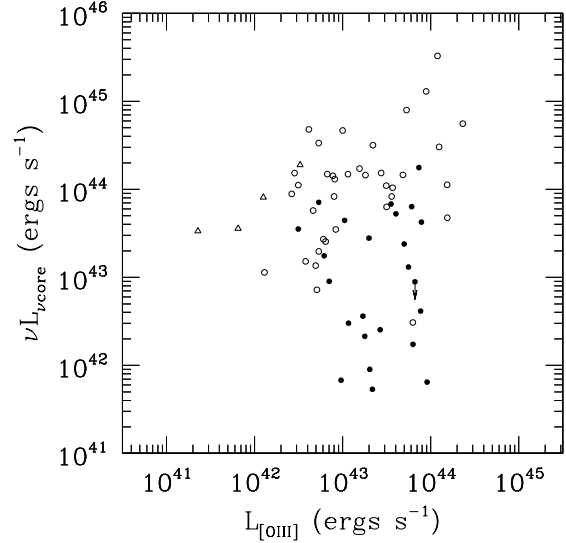,width=8.0cm,height=8.0cm}}
\caption{The radio core and narrow line [O\,{\sc iii}] luminosity relation
(Symbols as in Fig. 1).
}
\end{figure}

\begin{table*}
\begin{minipage}{150mm}
  \caption{Radio and line data of the sample.}
  \begin{tabular}{ccccllrll}\hline 
Source & Class. & z &  f$_{core}$(mJy) & f$_{ext}$(mJy) & $R$ & Ref.
& log f$_{[\rm O{\rm III }]}$ & Ref.   \\
(1) & (2) & (3) & (4) & (5) &(6) & (7) & (8) & (9) \\ 
\hline
 0014+813  &Q  &3.384  &  162.2  &    3.3  &   49.15  &K86  &            &  \\ 
 0056$-$001  &Q  &0.717  &  773.6  & $<$154.4  &  $>$5.01  &BM87  &  -13.96  &JB91  \\ 
 0106+013  &Q  &2.107  & 1560.9  &  196.5  &    7.94  &BM87  &            &  \\ 
 0112$-$017  &Q  &1.381  &  425.9  &   42.59  &   10.0  & BM87 &           &  \\ 
 0119+041  &Q  &0.637  &  521.5  &   32.9  &   15.85  &  BM87 &  -14.68  &JB91  \\   
 0133+207  &Q  &0.425  &   55.15  &  551.5  &    0.1  & BM87 &  -13.54  & JB91 \\ 
 0134+329  &Q  &0.367  &  389.1  & 4898.1  &    0.08  & BM87 &  -13.08  & JB91 \\ 
 0135$-$247  &Q  &0.831  &  530.5  &   33.47  &   15.85  &BM87 &  -14.03  & JB91 \\ 
 0159$-$117  &Q  &0.669  &  515.3  &   36.0  &   14.31  & WB86 &  -13.89  & T93 \\ 
 0212+735  &Q  &2.367  &  667.1  &    1.5  &  444.7  &NH90  &         &  \\ 
 0229+131  &Q  &2.065  &  660.4  &   33.1  &   19.95  &BM87  &         &  \\ 
 0234+285  &Q  &1.210  &  975.4  &    9.75  &  100.0  &BM87  &         &  \\ 
 0235+164  &BL  &0.940  &  501.0  &   10.3  &   48.64  &Mu93  &         &  \\ 
 0237$-$233  &Q  &2.224  & 1866.6  &    7.5  &  248.9  & NH90 &  -14.47  &B94  \\ 
 0248+430  &Q  &1.316  &  317.6  &    4.0  &   79.4  &BP86 &         &  \\ 
 0256+075  &Q  &0.893  &  291.6  &   12.8  &   22.78  &Mu93  &         &  \\ 
 0336$-$019  &Q  &0.852  & 1572.5  &   49.73  &   31.62  &BM87 &         &  \\ 
 0400+258  &Q  &2.109  &  391.0  &    3.91  &  100.0  &BM87  &         &  \\ 
 0403$-$132  &Q  &0.571  & 1715.1  &  342.2  &    5.01  & BM87 &  -14.41  & M96 \\ 
 0405$-$123  &Q  &0.574  &  482.8  & 1030.0  &    0.47  & M93 &  -13.35  & M96 \\ 
 0420$-$014  &Q  &0.915  & 2068.4  &    8.23  &  251.3  & BM87 &         &  \\ 
 0440$-$003  &Q  &0.844  &  641.5  &   32.15  &   19.95  & BM87 &  -14.70  & JB91 \\ 
 0458$-$020  &Q  &2.286  &  610.2  &  121.8  &    5.01  & BM87 &         &  \\ 
 0518+165  &Q  &0.759  &  422.4  & 3355.6  &    0.13  & BM87 &  -13.96  &JB91  \\ 
 0528$-$250  &Q  &2.765  &  268.06  &    6.73  &   39.83  &BM87  &         &  \\ 
 0537$-$441  &BL  &0.896  & 2185.5  &   10.95  &  199.6  &BM87  &         &  \\ 
 0537$-$286  &Q  &3.119  &  289.3  &    3.64  &   79.48  & BM87 &         &  \\ 
 0538+498  &Q  &0.545  &  304.5  & 7648.7  &    0.04  &BM87  &  -13.50  & L96 \\ 
 0539$-$057  &Q  &0.839  &  761.3  &  100.0  &    7.61  &U81  &  -15.15  &SK93  \\ 
 0602$-$319  &Q  &0.452  &   68.9  & 1000.0  &    0.07  &U81 &  -13.79  & R84 \\ 
 0605$-$085  &Q  &0.872  & 1467.0  &   51.4  &   28.54  &BP86  &         &  \\ 
 0607$-$157  &Q  &0.324  &  517.8  &  $<$51.78  &   $>$10.0  &BM87  &         &  \\ 
 0637$-$752  &Q  &0.654  & 2599.7  & 1880.0  &    1.38  & WB86 &  -13.27  & W99 \\ 
 0642+449  &Q  &3.396  &  104.4  &    1.8  &   58.0  &NH90 &         &  \\ 
 0711+356  &Q  &1.620  &  588.9  &    1.38  &  426.7  &Mu93 &         &  \\ 
 0723+679  &Q  &0.846  &  216.7  &  400.0  &    0.54  & U81 &  -14.59  & L96 \\ 
 0736+017  &Q  &0.191  & 1331.7  &   10.58  &  125.9  &BM87 &  -14.12  &T93  \\ 
 0738+313  &Q  &0.631  & 1321.3  &   21.3  &   62.03  &Mu93  &  -14.44  & JB91 \\ 
 0740+380  &Q  &1.063  &   $<$8.76  &  220.0  & $<$0.04  &BM87 &         &  \\ 
 0804+499  &Q  &1.433  &  264.3  &    1.74  &  151.9  & Mu93 &         &  \\ 
 0809+483  &Q  &0.871  &    2.96  & 4695.2  &  0.0006  &BM87 &  -13.68  & L96 \\ 
 0814+425  &BL  &0.258  & 1248.8  &   21.5  &   58.08  &Mu93  &         &  \\ 
 0820+225  &BL  &0.951  &  823.2  &  197.7  &    4.16  &Mu93  &         &  \\ 
 0823+033  &BL  &0.506  &  879.8  &    1.34  &  656.6  &Mu93  &         &  \\ 
 0825$-$202  &Q  &0.822  &  274.4  &  600.0  &    0.46  & U81 & -13.98  &d94  \\ 
 0834$-$201  &Q  &2.752  &  563.7  &    3.3  &  170.8  &NH90  &         &  \\ 
 0836+710  &Q  &2.172  & 1180.0  &   33.78  &   34.93  & Mu93 &         &  \\ 
 0838+133  &Q  &0.684  &  275.5  &  871.1  &    0.32  & BM87 &  -14.91  & M96 \\ 
 0842$-$754  &Q  &0.524  &  387.1  &  777.6  &    0.5  & M93 &  -13.86  & T93 \\ 
 0850+581  &Q  &1.322  &  123.8  &  115.5  &    1.07  &BP86  &         &  \\ 
 0851+202  &BL  &0.306  & 1562.0  &    2.0  &  781.0  &K92  &  -14.85  & S89 \\ 
 0858$-$279  &Q  &2.152  &  507.6  &  100.0  &    5.08  & U81 &         &  \\ 
 0859$-$140  &Q  &1.339  & 1202.0  &   52.6  &   22.85  & BP86 &         &  \\ 
 0859+470  &Q  &1.462  &  668.2  &  104.0  &    6.43  &Mu93  &         &  \\ 
 0906+015  &Q  &1.018  &  781.3  &   12.38  &   63.11  & BM87 &         &  \\ 
 0906+430  &Q  &0.668  &  583.2  &  734.2  &    0.79  &BM87  &  -14.66  &L96  \\ 
 0917+624  &Q  &1.446  &  455.8  &    2.1  &  217.1  &Mu93  &         &  \\ 
 0923+392  &Q  &0.698  & 4149.2  &  261.8  &   15.85  &BM87  &  -13.06  & L96 \\ 
 0945+408  &Q  &1.252  &  545.3  &   31.2  &   17.48  & Mu93 &         &  \\ 
 0953+254  &Q  &0.712  &  787.4  &    6.25  &  125.98  & BM87 &         &  \\ 
\hline
\end{tabular}
\end{minipage}
\end{table*}

\begin{table*}
 \begin{minipage}{150mm}
  \contcaption{Radio and line data of the sample.}
  \begin{tabular}{ccccllrll}
    &     &     &     &     &    &     &     &     \\    \hline
 0954+556  &Q  &0.901  & 1350.9  &  125.0  &   10.81  &Mu93  &  -14.33  & L96 \\ 
 0954+658  &BL  &0.367  &  994.9  &    9.0  &  110.5  &K92  & -15.47 &L96  \\ 
 1007+417  &Q  &0.6123  &  173.7  &  430.0  &    0.4  & WB86 &  -14.51  &M96  \\ 
 1040+123  &Q  &1.029  &  801.9  &  386.7  &    2.07  &Mu93  &         &  \\ 
 1045$-$188  &Q  &0.595  &  268.6  &  188.0  &    1.43  &BP86  &  -14.47  & S93 \\ 
 1055+018  &Q  &0.892  & 1481.5  &   69.1  &   21.44  &BP86  &         &  \\ 
 1100+772  &Q  &0.3115  &   89.75  &  712.9  &    0.13  & BM87 &  -13.43  &M96  \\ 
 1111+408  &Q  &0.734  &   11.55  &  729.1  &    0.02  & BM87 &  -13.68  &JB91  \\ 
 1127$-$145  &Q  &1.187  & 2051.6  &   40.93  &   50.12  & BM87 &         &  \\ 
 1136$-$135  &Q  &0.554  & $<$109.4  & 1730.0  &   $<$0.06  &WB86  &  -13.39  &T93  \\ 
 1137+660  &Q  &0.6563  &   76.74  &  966.04  &    0.08  &BM87  &  -14.52  &M96  \\ 
 1148$-$001  &Q  &1.982  &  927.2  &   15.4  &   60.21  &BP86  &         &  \\ 
 1150+497  &Q  &0.334  &  261.6  &   88.0  &    2.97  &S98  &  -14.03  &SM87  \\ 
 1226+023  &Q  &0.158  &25254.0  & 4002.6  &    6.31  & BM87 &  -12.38  & W99 \\ 
 1229$-$021  &Q  &1.045  &  500.7  &  145.8  &    3.43  &H83 &         &  \\ 
 1237$-$101  &Q  &0.753  &  542.7  &   34.2  &   15.87  &BM87  &         &  \\ 
 1250+568  &Q  &0.321  &   21.1  & 1055.3  &    0.02  & BM87 &  -13.37  & JB91 \\ 
 1253$-$055  &Q  &0.536  & 6326.0  &  502.5  &   12.59  &BM87  &  -14.56  & M96 \\ 
 1258+404  &Q  &1.6656  &   36.76  &  351.8  &    0.1  &H83  &        &  \\ 
 1302$-$102  &Q  &0.286  &  763.6  &   51.68  &   14.78  &L94  &  -14.02  & M96 \\ 
 1308+326  &BL  &0.997  &  505.8  &   14.0  &   36.13  & K92 &         &  \\ 
 1328+307  &Q  &0.846  & 3887.9  &  125.0  &   31.1  & K90 &  -13.89  &GW94  \\ 
 1334$-$127  &Q  &0.539  & 1046.0  &   81.3  &   12.87  & BP86 &         &  \\ 
 1340+606  &Q  &0.961  &    2.29  &  362.6  &    0.01  & BM87 &         &  \\ 
 1354+195  &Q  &0.720  &  680.0  &  290.0  &    2.34  &WB86  &         &  \\ 
 1355$-$416  &Q  &0.313  &   28.2  & 1347.0  &    0.02  & M93 &  -13.70  &T93  \\ 
 1416+067  &Q  &1.439  &  119.9  & 1198.8  &    0.1  & BM87 &         &  \\ 
 1424$-$418  &Q  &1.522  &  713.7  &  300.0  &    2.38  & U81 &         &  \\ 
 1442+101  &Q  &3.5305  &  288.9  &   91.36  &    3.16  &BM87  &         &  \\ 
 1451$-$375  &Q  &0.314  & 1088.3  &  410.0  &    2.65  & WB86 &         &  \\ 
 1458+718  &Q  &0.905  &  268.02  & 2680.2  &    0.1  &BM87 &  -13.89  & L96 \\ 
 1504$-$166  &Q  &0.876  & 1462.7  &  $<$73.3  &  $>$19.95  &BM87  &           &  \\ 
 1510$-$089  &Q  &0.361  & 2550.1  &   80.64  &   31.62  &BM87  &  -13.91  &T93  \\ 
 1512+370  &Q  &0.371  &   87.5  &  430.0  &    0.2  &WB86  &  -13.77  &M96  \\ 
 1532+016  &Q  &1.435  &  427.7  &    5.2  &   82.25  & BM87 &         &  \\ 
 1538+149  &BL  &0.605  &  828.7  &   66.3  &   12.5  &Mu93  &         &  \\ 
 1546+027  &Q  &0.412  &  812.3  &    6.17  &  131.7  &Mu93  & -14.01& B89  \\ 
 1555+001  &Q  &1.770  &  330.47  &  $<$33.0  &  $>$10.01  &BM87  &         &  \\ 
 1611+343  &Q  &1.401  & 1027.2  &   40.89  &   25.12  &BM87  &         &  \\ 
 1622$-$253  &Q  &0.786  & 1735.7  &  300.0  &    5.79  & U81 &         &  \\ 
 1633+382  &Q  &1.814  &  772.18  &    9.72  &   79.44  &BM87  &         &  \\ 
 1637+574  &Q  &0.750  &  976.0  &  110.0  &    8.87  &K90  &  -14.06  & M96 \\ 
 1638+398  &Q  &1.666  &  282.1  &    8.0  &   35.26  &K90  &         &  \\ 
 1641+399  &Q  &0.594  & 4923.1  &  155.7  &   31.62  &BM87 &  -14.28  & L96 \\ 
 1642+690  &Q  &0.751  &  708.7  &   74.46  &    9.52  &S98  &  -15.00  &L96  \\ 
 1704+608  &Q  &0.371  &  120.0  &  953.1  &    0.13  &BM87  &  -12.95  & W99 \\ 
 1721+343  &Q  &0.206  &  306.8  &  270.0  &    1.14  & WB86 &  -12.50  & W99 \\ 
 1725+044  &Q  &0.293  &  650.5  &    5.17  &  125.8  & BM87 &  -13.93  & R84 \\ 
 1739+522  &Q  &1.379  &  339.7  &    6.78  &   50.1  &BM87  &         &  \\ 
 1803+784  &BL  &0.684  & 1472.7  &    3.0  &  490.9  &K92  &  -14.89  &L96  \\ 
 1823+568  &BL  &0.664  &  673.1  &  198.0  &    3.4  &K92  &  -15.28 & L96 \\ 
 1828+487  &Q  &0.691  & 1345.8  & 4255.7  &    0.32  &BM87 &  -13.55  &L96  \\ 
 1830+285  &Q  &0.594  &  371.4  &  127.5  &    2.91  & H92 &         &  \\ 
 1928+738  &Q  &0.302  & 2132.0  &   71.0  &   30.03  &K90  &  -12.46  & L96 \\ 
 1954$-$388  &Q  &0.626  & 1218.0  &    9.2  &  132.4  & M97 &        &  \\ 
 1954+513  &Q  &1.230  &  652.0  &  144.0  &    4.53  &K90  &       &  \\ 
 2029+121  &BL  &1.215  &  323.0  &   15.1  &   21.39  &BP86  &        &  \\ 
 2126$-$158  &Q  &3.266  &  134.6  &    8.4  &   16.02  &NH90  &       &  \\ 
 2128$-$123  &Q  &0.501  & 1266.0  &   46.0  &   27.52  &M97 &  -13.56  & T93 \\ 
 2134+004  &Q  &1.936  & 1641.0  &    2.16  &  759.7  &Mu93  &        &  \\ 
 2135$-$147  &Q  &0.200  &   95.7  &  479.6  &    0.2  &BM87 &  -12.97  & W99 \\ 
 2136+141  &Q  &2.427  &  365.6  &    0.3  & 1218.7  &NH90  &        &  \\ 
 2145+067  &Q  &0.990  & 1334.7  &    5.25  &  254.3  &Mu93  &        &  \\ 
 2155$-$152  &Q  &0.672  &  717.7  &  108.4  &    6.62  & W84 &  -14.97  & S89 \\ 
                                   \hline

 \end{tabular}
\end{minipage}
\end{table*}

\begin{table*}
 \begin{minipage}{150mm}
  \contcaption{Radio and line data of the sample.}
  \begin{tabular}{ccccllrll}
    &     &     &     &     &    &     &     &     \\   \hline
 2201+315  &Q  &0.298  & 1249.2  &  788.2  &    1.58  &BM87  &  -13.86  & M96 \\ 
 2203$-$188  &Q  &0.619  & 1402.0  & 1348.0  &    1.04  &M97  & -14.06  & S89 \\ 
 2209+080  &Q  &0.484  &  158.4  &  303.0  &    0.52  & H92 &         &  \\ 
 2216$-$038  &Q  &0.901  &  857.5  &   68.11  &   12.59  &BM87  &         &  \\ 
 2223$-$052  &Q  &1.404  & 2037.2  &   64.42  &   31.62  &BM87  &         &  \\ 
 2223+210  &Q  &1.949  &  598.8  &   15.4  &   38.88  &Mu93  &       &  \\ 
 2230+114  &Q  &1.037  & 1867.7  &   74.35  &   25.12  & BM87 &        &  \\ 
 2234+282  &Q  &0.795  & 1880.4  &    7.49  &  251.19  &BM87  &  -14.82  &JB91  \\ 
 2240$-$260  &BL  &0.774  &  475.8  &   90.6  &    5.24  &C99  &        &  \\ 
 2243$-$123  &Q  &0.630  & 1601.0  &   59.8  &   26.77  &M97  &  -14.14  &T93  \\ 
 2247+140  &Q  &0.237  & 1469.0  &   34.3  &   42.83  & BP86 &  -13.70  &GW94  \\ 
 2251+158  &Q  &0.859  & 6128.3  &  386.7  &   15.85  &BM87  &  -13.68  &JB91  \\ 
 2318+049  &Q  &0.623  &  276.0  &    9.0  &   30.66  & BP86 &          &  \\ 
 2319+272  &Q  &1.253  &  374.6  &   43.7  &    8.57  &BP86  &         &  \\ 
 2328+107  &Q  &1.489  &  427.88  &    8.2  &   52.18  &Mu93  &        &  \\ 
 2344+092  &Q  &0.673  &  907.25  &    9.07  &  100.0  & BM87 &  -13.21  &JB91  \\ 
 2345$-$167  &Q  &0.576  & 1678.9  &   66.84  &   25.12  &BM87 &  -14.19  & JB91 \\ 
                                                 \hline
\end{tabular}

\medskip

Notes for the table 1. Q: quasars; BL: BL Lac objects.\\
Column (1): IAU source name. Column (2): classification of the source.
Column (3): redshift. Column (4): core flux density in the rest frame
of the source. Column (5): extended flux density in the rest frame
of the source. Column (6): ratio $R$ of the core to extended
emission. Column (7): references for the radio emission. 
Column (8): flux of ${[\rm O{\rm III
}]}$ (in ergs.~s$^{-1}$~cm$^{-2}$).
Column (9): references for the line emission. \\

\vskip 1mm
References:

 B89: Baldwin et al. (1989).
 B94: Baker et al. (1994). 
BM87: Browne \& Murphy (1987). 
BP86: Browne \& Perley (1986). 
C99: Cassaro et al. (1999). 
 d94: di Serego Alighieri et al. (1994). 
 GW94: Gelderman \& Whittle (1994).
H83: Hintzen et al. (1983).  
H92: Hooimeyer et al. (1992). 
 JB91: Jackson \& Browne (1991).
K86: K\"uhr et al. (1986). 
K90: Kollgaard et al. (1990). 
K92: Kollgaard et al. (1992). 
L94: Lister et al. (1994). 
 L96: Lawrence et al. (1996).
M93: Morganti et al. (1993). 
 M96: Marziani et al. (1996). 
M97: Morganti et al. (1997). 
Mu93: Murphy et al. (1993). 
NH90: Neff \& Hutchings (1990). 
 R84: Rudy (1984). 
 S89: Stickel et al. (1989).
 S93: Stickel et al. (1993). 
S98: Saikia et al. (1998). 
 SK93: Stickel \& K\"uhr (1993). 
 SM87: Stockton \& MacKenty (1987). 
 T93: Tadhunter et al. (1993). 
U81: Ulvestad et al. (1981). 
W84: Wardle et al. (1984). 
 W99: Wilkes et al. (1999). 
WB86: Wills \& Browne (1986). 
\end{minipage}

\end{table*}

\section{Results}

We present the core and extended radio luminosity at 5 GHz in the rest
frame of the source as functions of redshift $z$ for the sample in
Figs. 1 and 2. We note that the extended radio luminosity provides
wider dispersion  and show weaker dependence
of redshift than the core luminosity. This is due to the fact that
present sample has a limit
on total radio flux density, not on extended radio flux 
density. We plot
the relation between the extended radio luminosity and total broad-line
luminosity in Fig. 3 and find a significant correlation between them at
$>$ 99.9 per cent confidence (Spearman correlation coefficient
$\rho$). 
It is well known that the correlation between luminosities may be caused by common
redshift dependence. We therefore perform a statistic analysis on the
sources in the restricted redshift range $0.5 <z<1.5 $. For this subsample of
sources, we check the correlation between luminosity and redshift,
and no correlation between the extended radio luminosity and
redshift is found (at 15  per cent confidence, also see Fig. 2), 
while a significant 
correlation is still present at 98 per cent confidence between the extended
luminosity and  total broad-line luminosity.  
The correlation between the extended
radio flux and the broad-line flux confirms these analyses 
on luminosities (Fig. 4).
We also use the Spearman partial rank correlation method (Macklin
1982) to check the correlations. The statistic results are listed in
Table 2. We note that there are significant correlations between 
$\nu L_{\nu{\rm ext}}$ and $L_{\rm line}$ independent of $z$. There
is almost no correlation between  $\nu L_{\nu{\rm ext}}$ and $z$ 
independent of $L_{\rm line}$ either for the whole sample or 
the redshift restricted subsample.

The relation between the core luminosity and total broad-line luminosity
is given in Fig. 5.  
A strong correlation is present for the CDQs
in the sample, and we can see that the behaviours of CDQs and
LDQs are rather different.  Similar phenomena can be
seen in the relation between fluxes (see Fig. 6). We re-examine such 
relations between radio and narrow
line [O\,{\sc iii}] emission in Figs. 7 and 8, and similar results are
present. The statistic results are almost same if those sources only
with an upper limit on their core or extended luminosity are ruled out
in the analyses. The BL Lac objects in present sample show nothing
speacial except for their weak line emission.

\begin{table*}
 \begin{minipage}{150mm}
  \caption{Spearman partial rank correlation analysis of the
  correlations present between $\nu L_{\nu{\rm ext}}$, 
$L_{\rm line}$ and $z$. $r_{\rm AB}$ is the rank correlation
  coefficient of the two variables and $r_{\rm AB,C}$ the partial 
rank correlation coefficient. The significance of the partial 
rank correlation is equivalent to the deviation from a unit 
variance normal distribution if there is no correlation present. 
}
  \begin{tabular}{lrcccc}\hline
Sample  & N & Correlated variables: A,B & $r_{\rm AB}$ & $r_{\rm AB,C}$
  & significance \\ \hline
All & 141 & $\nu L_{\nu{\rm ext}}$, $L_{\rm line}$ & 0.303 & 0.240 &
  2.868\\
    &   & $z$, $L_{\rm line}$ & 0.438 & 0.402 & 4.983\\
    &   & $\nu L_{\nu{\rm ext}}$, $z$ & 0.210 & 0.090 & 1.058 \\
within  & 90 & $\nu L_{\nu{\rm ext}}$, $L_{\rm line}$ &
  0.246 & 0.255 &2.417\\
$0.5<z<1.5$ &   & $z$, $L_{\rm line}$ & 0.199 & 0.210 & 1.979\\
    &   & $\nu L_{\nu{\rm ext}}$, $z$ & $-$0.021~~ & $-$0.073~~
    & $-$0.679~~~\\ \hline
 \end{tabular}
\end{minipage}
\end{table*}

\section{Discussion}

The radio emission for steep-spectrum quasars is believed to
be unbeamed emission from the lobes. 
The dominant influence on the radio luminosity is $Q_{\rm jet}$, and not the
large-scale radio source environment (Serjeant et al. 1998). 
Therefore the radio emission can be a measure of jet power $Q_{\rm jet}$. 
The core emission in flat-spectrum quasars is strongly 
beamed to us, but the extended emission is not. 
If the difference between flat and steep-spectrum quasars is caused 
by the different angles between the jet orientation and the line of
sight, then one can take the
extended radio emission from flat-spectrum quasars as a tracer of
jet power, as that for steep-spectrum quasars.  
The optical continuum  is a good indicator of the disc surrounding
a black hole for steep-spectrum quasars, 
since relativistic beaming does not affect the optical
continuum in these sources.
Thus, the radio-optical correlation gives evidence of
a disc-jet link (Serjeant et al. 1998).
For flat-spectrum quasars, the optical continuum may be contaminated
by the beamed synchrotron emission from jets, which prevents us
from using it as an indicator of accretion power. The line emission
(narrow or broad lines) can alternatively be an indicator of accretion
power for both flat or steep-spectrum quasars. 

We present correlations between the extended radio and broad-line
emission for a sample of radio-loud quasars, most of which are
CDQs (99 of 128, if we define CDQs as $R>1$). We find  significant 
correlations
between the line emission (broad or narrow line) and extended
radio emission. 
A significant intrinsic correlation is also found
between the extended radio luminosity and total broad-line
luminosity for a subsample in the retricted redshift range, while no 
correlation is present between the extended radio luminosity and 
redshift for this subsample. The Spearman partial rank correlation 
analyses confirm this result. We therefore believe the correlation found
here is an intrinsic one which indicates a physical link between
jets and accretion processes. 
The CDQs  and LDQs follow the same
statistic behaviour, while the LDQs have relatively
high extended radio luminosity than their counterparts.  
The correlations between line and radio core emission exist for the
CDQs, while the LDQs show rather
different behaviours.

Another method to infer jet power is through the determination of
the physical quantities of jets, such as, the Lorentz factor, the size of the
jet and the density of electrons in jets, from radio and X-ray
observations(Celotti \& Fabian 1993). The radio emission from
CDQs is beamed to us. The Doppler factor of the jet
is the dominant influence on the observed radio luminosity, and radio
luminosity is also determined by the size of the jet, the electron
density, and the magnetic energy density in the jet. The core radio
luminosity for CDQs can therefore reflect the jet power for CDQs 
to some extent, though the uncertainty exists on the angle between the
jet axis orientation and line of sight. That might be the reason why we can 
also find a correlation between radio core emission and total
broad-line emission for the CDQs in our sample. A similar correlation 
is present between the core radio emission and the narrow line 
[O\,{\sc iii}] emission, which seems to rule out the possibility that 
the correlation between the core emission and broad-line emission is 
caused by the orientation effect of broad-line region, i.e., a jet
close to the line of sight may imply an unobstructed view of the broad-line
region. It is worth pointing  out that the core radio
emisson is not a good tracer for jet power of LDQs (see Figs. 5--8). 
Instead, the extended radio emission can be taken as 
a tracer for jet power both for CDQs and LDQs. The jet power 
can be derived from their extended radio luminosity based on 
some assumptions (see detailed discussion in Willott et al. 1999).

Similar correlation analyses are
performed between radio and narrow line [O\,{\sc iii}]
emission. Although the sample is reduced to a rather smaller one due
to the lack of the narrow line data mainly caused by the redshift
requirement on observations of [O\,{\sc iii}], the similar results are
present as that for the total broad-line(see Figs. 7 and 8).

\section*{Acknowledgments}
We thank Peter Scheuer and the anonymous referee for their helpful 
comments and suggestions.  
The support from the NSFC and Pandeng project is gratefully acknowledged. 
This research has made use
of the NASA/IPAC Extragalactic Database (NED), which is operated by the Jet
Propulsion Laboratory, California Institute of Technology, under contract
with the National Aeronautic and Space Administration.

{}


\begin{thebibliography}{}


\bibitem{}Antonucci R.R.J., 1993, ARA\&A, 31, 473
\bibitem{}Baker A.C., Carswell R.F., Bailey J.A., Espey B.R., Smith M.G.,
    Ward M.J., 1994, MNRAS 270, 575(B94)
\bibitem{}Baldwin J.A., Wampler E.J., Gaskell C.M., 1989, ApJ, 338, 630(B89) 
\bibitem{}Barthel P.D., 1989, ApJ, 336, 606
\bibitem{}Baum S.A., Heckman T.M., 1989, ApJ, 336, 702
\bibitem{}Blandford R. D., Payne D. G., 1982, MNRAS, 199, 883
\bibitem{}Blandford R. D., Znajek R. L., 1977, MNRAS, 179, 433
\bibitem{}Browne I.W.A., Murphy D.W., 1987, MNRAS, 226, 601(BM87)
\bibitem{}Browne I.W.A., Perley R.A., 1986, MNRAS, 222, 149(BP86)
\bibitem{}Cao X., Jiang D.R., 1999, MNRAS, 307, 802
\bibitem{}Cassaro P., Stanghellini C., Bondi M., Dallacasa D., Ceca
R.D., Zappala R.A., 1999, A\&AS, 139, 601(C99)
\bibitem{}Celotti A., Fabian A.C., 1993, MNRAS, 264, 228
\bibitem{}di Serego Alighieri S., Danziger I.J., Morganti R., Tadhunter C.N., 
  1994, MNRAS, 269, 998(d94) 
\bibitem{}Dunlop J.S., Peacock J.A., 1993, MNRAS, 263, 93
\bibitem{}Falcke H., Biermann P., 1995, A\&A, 293, 665
\bibitem{}Francis P.J., Hewett P.C., Foltz C.B., Chaffee F.H., Weymann R.J.,
   Morris S.L., 1991, ApJ, 373, 465
\bibitem{}Gelderman R., Whittle M., 1994, ApJS, 91, 491(GW94) 
\bibitem{}Hintzen P., Ulvestad J., Owen F., 1983, AJ, 88, 709(H83)
\bibitem{}Hooimeyer J.R.A., Schilizzi R.T., Miley G.K., Bathel P.D.,
1992, A\&A, 261, 25(H92) 
\bibitem{}Jackson N., Browne I.W.A., 1991, MNRAS, 250, 414(JB91)
\bibitem{}Koide S., Shibata K., Kudoh T., 1999, ApJ, 522, 727 
\bibitem{}Kollgaard R.I., Wardle J.F.C., Roberts D.H., 1990, AJ, 100,
1057(K90)
\bibitem{}Kollgaard R.I., Wardle J.F.C., Roberts D.H., Gabuzda D.C.,  1990, AJ, 104,
1687(K92)
\bibitem{}K\"uhr H., Stocke J.T., Strittmatter P.A., Bartel N., 1986,
ApJ, 302, 52(K86)
\bibitem{}Lawrence C.R., Zucker J.R., Readhead A.C.S., Unwin S.C.,
   Pearson T.J. et al., 1996, ApJS, 107, 541(L96)
\bibitem{}Lister M.L., Gower A.C., Hutchings J.B., 1994, AJ, 108,
821(L94)
\bibitem{}Macklin J.T., 1982, MNRAS, 199, 1119
\bibitem{}Marziani P., Sulentic J.W., Dultzin-Hacyan D., Calvani M., Mole
   M., 1996, ApJS, 104, 37(M96)
\bibitem{}McCarthy P.J., 1993, ARA\&A, 31, 639 
\bibitem{}Morganti R., Killeen N.E.B., Tadhunter C.N., 1993, MNRAS,
263, 1023(M93)
\bibitem{}Morganti R., Oosterloo T.A., Reynolds J.E., Tadhunter C.N.,
Migenes V., 1997, MNRAS, 284, 541(M97)
\bibitem{}Murphy D.W., Browne I.W.A., Perley R.A., 1993, MNRAS, 264,
298(Mu93)
\bibitem{}Neff S.G., Hutchings J.B., 1990, AJ, 100, 1441(NH90)
\bibitem{}Rawlings S. G., Saunders R. D. E., 1991, Nat, 349, 138
\bibitem{}Rawlings S. G., Saunders R. D. E., Eales S.A., Mackay C.D.,
1989, MNRAS, 240, 701
\bibitem{}Rudy R.J., 1984, ApJ, 284, 33(R84)
\bibitem{}Saikia D.J., Holmes G.F., Kulkarm A.R., Salter C.J.,
Garrington S.T., 1998, MNRAS, 298, 877(S98)
\bibitem{}Scheuer P.A.G., 1987, in Zensus J.A., Pearson T.J., eds,
Superluminal Radio Sources. Cambridge Univ. Press, Cambridge, p. 331
\bibitem{}Serjeant S., Rawlings S., Maddox S.J., Baker J.C.,
   Clements D., Lacy M., Lilje P.B., 1998, MNRAS, 294, 494
\bibitem{}Stickel M., K\"uhr H., 1993, A\&AS, 101, 521(SK93)
\bibitem{}Stickel M., Fried W., K\"uhr H., 1989, A\&AS, 80, 103(S89)
\bibitem{}Stickel M., K\"uhr H., Fried W., 1993, A\&AS, 97, 483(S93)
\bibitem{}Stockton A., MacKenty J.W., 1987, ApJ, 316, 584(SM87) 
\bibitem{}Tadhunter C.N., Morganti R., Alighieri S. et al., 1993,
  MNRAS, 263, 999(T93) 
\bibitem{}Tadhunter C.N., Morganti R., Robinson A., Dickson R.,
  Villar-Martin M., Fosbury R.A.E., 1998, MNRAS, 298, 1035 
\bibitem{}Ulvestad J., Johnston K., Perley R., Fomalont E., 1981, AJ,
86, 1010(U81)
\bibitem{}Wardle J.F.C., Moore R.L., Angel J.R.P., 1984, ApJ, 279, 93(W84)
\bibitem{}Wilkes B.J., Kuraszkiewicz J., Green P.J., Mathur S.,
McDowell J.C., 1999, ApJ, 513, 76(W99)
\bibitem{}Willott C.J., Rawlings S., Blundell K.M., Lacy M., 1998,
MNRAS, 300, 625
\bibitem{}Willott C.J., Rawlings S., Blundell K.M., Lacy M., 1999,
MNRAS, 309, 1017
\bibitem{}Wills B.J., Browne I.W.A., 1986, ApJ, 302, 56(WB86)
\bibitem{}Xu C., Livio M., 1999, AJ, 118, 1169
\end{thebibliography}
\end{document}